\documentclass[3p,times]{elsarticle}

\usepackage{ecrc}

\volume{00}

\firstpage{1}

\journalname{Procedia Computer Science}

\runauth{}

\jid{procs}

\jnltitlelogo{Procedia Computer Science}

\CopyrightLine{2011}{Published by Elsevier Ltd.}

\usepackage{amssymb}

\usepackage[figuresright]{rotating}

\usepackage[figuresright]{rotating}

\newcommand\nn{\nonumber}
\newcommand\ba{\begin{eqnarray}}
\newcommand\ea{\end{eqnarray}}
\newcommand\alb{\begin{align}}
\newcommand\ale{\end{align}}
\newcommand\be{\begin{equation}}
\newcommand\ee{\end{equation}}

\begin{document}

\begin{frontmatter}
\dochead{}

 \title{A model for space and time-like proton (neutron) form factors}


\author[label1]{E.A.Kuraev} 
\author[label2]{A. Dbeyssi \corref{cor1}}
\author[label3,label2]{E.~Tomasi-Gustafsson}
\ead{etomasi@cea.fr}
\address[label1]{Joint Institute for Nuclear Research, Dubna, Russia}
\cortext[cor1]{Boursier du CNRS libanais/LNCSR Scholar}
\address[label2]{Univ Paris-Sud, CNRS/IN2P3, Institut de Physique Nucl\'eaire, UMR 8608, 91405 Orsay, France}
\address[label3]{CEA,IRFU,SPhN, Saclay, 91191 Gif-sur-Yvette, France}

\begin{abstract}
A model is suggested to interpret nucleon electromagnetic form factors both in space and time-like regions and to describe the existing data. It implies quarks to be colorless in the region of high intensity chromo-electromagnetic gluon field inside the nucleon. 
\end{abstract}


\end{frontmatter}


\section{Introduction}

Electromagnetic form factors (FFs) describe the internal structure of a hadron, its charge and magnetic distributions. Scattering ($e+p\to e+p$) and annihilation ($\bar p+p\leftrightarrow e^++e^-$) reactions, at the lowest order of perturbation theory (Born approximation) contain the same vertex $\gamma^* p p(\bar p)$, if one assumes the exchange of a virtual photon with square momentum $q^2=-Q^2$. Such vertex is parametrized in terms of FFs which are function of  $q^2$ only. Note that $q^2>0$ in the time-like (TL) region and $Q^2>0$ in space-like (SL) region. 

A simple correspondence exists between the electric FF and the charge, the magnetic FF and the magnetic moment of the nucleon, through the normalization at the photon point ($Q^2=0$). At $Q^2>0$ FFs are the Fourier transform of the charge and magnetic distributions, in non-relativistic approaches, and also in the Breit frame, the system where the energy of the virtual photon vanishes and its four momentum coincides with the three momentum. In another reference system (Laboratory or Center of Mass) and in the TL region, the relation between FFs and physical distributions is not straightforward and  requires model assumptions. Attempts to find this correspondence have been done in the SL region \cite{Miller:2003sa,Cloet:2008re}. 

Different nucleon models exist (for a review, see for example \cite{Perdrisat:2006hj}), but not all of them describe all four nucleon FFs and not all of them have the necessary analytical properties to be extended in TL region \cite{TomasiGustafsson:2005kc,TomasiGustafsson:2005sk}. 

The understanding of the existing data and the future plans of precise measurements on nucleon FFs motivate the necessity of new models of the nucleon at relatively large momentum transfer squared, which are able to explain the existing data in SL as well as in TL regions and to give useful predictions.

The purpose of this work is to suggest a qualitative model which applies to proton and neutron and which is able to describe data issued both from scattering and annihilation processes. The present model  gives an explanation for the recent experimental finding of the fast decreasing of the charge distribution at small distances inside the proton by the GEp collaboration \cite{Pu10} based on the polarization method \cite{Re67}. It suggests that a similar behavior should be found in TL region. Indeed, two experiments measured the ratio between the electric $G_E$ and the magnetic $G_M$ FFs. Although  the results are affected by large error bars, data seem inconsistent, suggesting a ratio larger than unity in Ref. \cite{Babar}, whereas it  decreases being always smaller than unity in Ref. \cite{Ba94}.

In Section 2 a possible scenario for the interaction is suggested. In Subsection 2.1 the general considerations are applied to the scattering region, and an effective FFs parametrization in SL region, is given. In Subsection 2.2 we generalize the definition of FFs to the annihilation channel. In Subsection 2.3 we apply our considerations to TL FFs. In Section 3 the suggested  parametrization is compared to the existing data. Section 4 contains the concluding remarks.

\section{General considerations}

Our considerations are derived in the Breit frame for SL region and in the center of mass system (CMS) in TL region. FFs will be derived as function of $q^2=-Q^2$, which is invariant. 

Let us do an analogy with gravitation. Newton showed that for a spherical symmetric distributed mass density, a point located at a distance $r<R_0$ from the center ($R_0$ is the radius of the object) is submitted to a gravitational force due only to the matter located inside the surface where the point is located. The gravitational interaction with the layer of matter situated at $r>R$ is fully compensated and its effect vanishes. The Coulomb force between the electron and the proton has a similar behavior as the gravitational force. The Newton-like theorem mentioned above, works only for the scalar part $\phi$ of the electromagnetic field, $A=(\phi,\vec A)$,  not for the vector $\vec A$, which is related to the magnetic field. If one applies these arguments to the electric and magnetic proton FFs, one finds that their ratio should be:
\be
\frac{\mu G_E}{G_M} \simeq \left (\frac{1}{rQ}\right )^3, ~Q> \frac{1}{r},
\ee
where $\mu$ is the proton anomalous magnetic moment ($\mu=2.79$ in units of Bohr magnetons). For example, a value of the internal distance $r\le 0.7$ fm, corresponds to a momentum $Q\ge 0.29$ GeV. The experimental data in this kinematical range exclude a $Q^{-3}$ dependence of this ratio, so that the description of the nucleon as a homogeneous electric sphere is not appropriate.

The general understanding of the proton structure is of a system of three valence quarks and of a neutral sea consisting of gluons and quark-antiquark pairs. It is usually assumed that the nucleon is an antisymmetric state of colored quarks, that one can write formally as ($i,j,k$ are the color index):
\be
|p>\sim\epsilon_{ijk}|u^i u^j d^k>, ~
|n>\sim\epsilon_{ijk}|u^i d^j d^k>.
\label{eq:eq1}
\ee
The main idea of this work is that the spatial center of the nucleon (proton and neutron) is electrically neutral, therefore the description (\ref{eq:eq1}) does not apply, starting at some value of $Q$ or for distances smaller than some radius. In other words, the  representation (\ref{eq:eq1}) fails inside the nucleon where the region of a strong gluonic field creates a gluonic condensate of clusters with a randomly oriented chromo-magnetic field \cite{Vainshtein:1982zc}. Let us take for the mean square of the intensity of the gluon field in vacuum ($G$ refers to the gluon field tensor $G_{\mu\nu}$):
\be 
<0|\frac{\alpha_s}{\pi}(G^a_{\mu\nu})^2|0> =\frac{\alpha_s}{\pi}G^2=0.012 \mbox{~GeV~}^4.
\label{eq:eq2}
\ee
Taking $\alpha_s/\pi\sim 0.1$, and neglecting the chromo-magnetic field, one can find that the strength of the chromo-electric field, $E$, is: 
\be
G^2=2(E^2-B^2)\approx 2 E^2 = 0.012~\frac{\pi}{\alpha_s} ~\mbox{~GeV}^4,\mbox{~i.e.},~ E\simeq~0.245~~\mbox{~GeV}^2.
\label{eq:eq2a}
\ee
At smaller distances the gluonic field as well as the number of gluons increases. Therefore, the strength of the chromo-electric field increases too. Our main assumption consists in the statement that in the very internal region of strong chromo-electromagnetic field the color quantum number of quarks does not play any role, due to stochastic averaging. Let us formalize this property in the following way:
\be
<G|u^i u^j|G>\sim \delta_{ij},
\label{eq:eq3}
\ee
(for the neutron, the same property applies, where the $u$ quark is replaced by the $d$ quark). When the color quantum number of quarks of the same flavor vanishes, then, due to Pauli principle, $uu$ (or $dd$) quarks move away, outside the internal region of the proton (or the neutron). The third quark is attracted by one of the identical quarks and forms a compact diquark. The strength of the gluonic field decreases and the color state of quarks is restored. Let us estimate the distance (or transferred momentum) where this may happen. The creation of a quark-diquark dipole system occurs when the attraction force exceeds the stochastic force of the gluon field:
\be
\displaystyle\frac{Q_q^2 e^2}{r_0^2}> e|Q_q|~ E.
\ee
where $eQ_q$ is the charge of the isolated quark which is $u(-1/3)$ or $d(2/3)$ in case of proton(neutron) and $r_0$ is the distance between the quark and the diquark system. Taking into account the relation between the conjugated variables:
$ r_0p_0\approx  1$ (assuming $\hbar =c=1$), one obtains for proton:
\be
~p_0=\sqrt{\displaystyle\frac{E}{e|Q_q|}}=1.1 ~\mbox{GeV}.
\ee
The minimal distance where the quark-diquark picture appears is $r_0= 0.22$ (0.31) fm in the proton (neutron), which corresponds to $p_0^2\approx 1.21 (2.43)$  GeV$^2$. Therefore our considerations apply in a different range for proton and neutron, at least in the SL region. Moreover they apply to the scalar part of the field, leaving unchanged the predictions from quark counting rules for the magnetic FF.

Quark counting rules apply to the vector part of the interaction, and are derived from the interaction of the virtual photon with three independent quarks. The corresponding behavior of FFs for the nucleon is \cite{Ma73,Br73}: 
\ba
G_M^{(p,n)}(Q^2)&=&\mu G_E(Q^2),\nn\\
G_E^{(p,n)}(Q^2)&=&G_D(Q^2)=\left [ 1+Q^2/(0.71 \mbox{~GeV}^2)\right ]^{-2},
\label{eq:GMSL}
\ea
where $Q^2$ is expressed in GeV$^2$ and FFs are normalized respectively to the nucleon charge and magnetic moment: $G_{E}^{(p,n)}(0)=(1,0)$, $G_{M}^{(p,n)}(0)=\mu_{p,n}$.

\subsection{The scattering channel}

Let us consider the scattering channel. The additional suppression mechanism of the electric FF is provided by the 'central' region of the hadron. A similar effect to the screening of a charge in plasma may take place. The scalar part of the electromagnetic field $\phi$ obeys to the equation: 
\be
\Delta\phi=-4\pi e\sum Z_i n_i,~ n_i=n_{i0}exp{\left[-\displaystyle\frac{Z_ie\phi}{kT}\right]},
\label{eq:eq4}
\ee
where $Z_i$ ($n_i$) are the charges (numbers) of current quarks and antiquarks in the neutral charge region, $k$ is the Boltzmann constant, and $T$ is the effective temperature of the hot plasma. Expanding the Boltzmann exponent and using the neutrality condition: 
$\sum Z_i n_{i0}=0$ we obtain:
\be
\Delta\phi- \chi^2\phi=0,~ \phi=\displaystyle\frac {e^{-\chi r}}{r},
~\chi^2=\displaystyle\frac{ 4\pi e^2 Z_i^2n_{i0}}{kT}.
\label{eq:eq5}
\ee
The distribution in momentum space, applying the Fourier transform, shows that the central region provides an additional suppression for the electric FF:
\be
G_E(Q^2)=\displaystyle\frac {G_M(Q^2)}{\mu }\left(1+{Q^2}/{q_1^2} \right )^{-1}
\label{eq:GESL}
\ee
where $q_1(\equiv \chi)$ should be considered a fitting parameter, different in principle for proton and neutrons. This explains the observed monopole decreasing for the proton FF ratio in SL region.

\subsection{\it  Generalized Form Factors}

We generalize the definition of FFs as:
\be
F(q^2)=\int_{\cal D} d^4x e^{iq_{\mu}x^{\mu}}\rho(x),~q_{\mu}x^{\mu}=q_0t-\vec q\cdot\vec x
\label{eq:eq5b}
\ee
where $\rho(x)=\rho(\vec x,t) $ is the space-time distribution of the electric charge in a space-time volume ${\cal D}$. In the scattering channel, $e+p\to e+p$ and in the Breit frame, we recover the usual definition of FFs:
\be
F(q^2)=\delta(q_0) F(Q^2),~Q^2=-(q_0^2-\vec q^2)>0,
\ee
where zero energy transfer is implied. 

In the annihilation channel, in the center of mass reference frame where the three-dimensional transferred momentum vanishes, $q_0 = \sqrt{q^2}$, and we have:
\be
F(q^2)=\int_{\cal D} dt e^{i\sqrt{q^2} t }\int d^3 \vec r \rho(\vec r,t)=
\int_{\cal D}  dt e^{i\sqrt{q^2} t } {\cal Q}(t),
\label{eq:eq14}
\ee
where ${\cal Q}(t)$ is a scalar quantity which can be interpreted as the time evolution of the charge distribution in the physical domain ${\cal D}$.

\subsection{ The annihilation channel}

Let us consider the annihilation channel and the process $e^++e^-\to \gamma^*(q)\to p+\bar p$. The mechanism of $\bar p p$ creation in $e^+e^-$ annihilation through a spin one intermediate state $^3S_1=<0|J^{\mu}|p\bar p>$ is essentially different from elastic $ep$ scattering, although there is a connection of the underlying physics in terms of FFs.

 We can describe this process in three steps.

1) Above the physical threshold, $q^2\ge 4M_p^2$, the vacuum state transfers all the energy released by the electron-positron annihilation to a S-wave state with total spin 1 and four momentum $q=(\sqrt{q^2},0,0,0)$. This state of matter consists in at least six massless valence  quarks, a set of gluons and a sea of current $q\bar q$ quarks, with total energy $q_0>2M_p$, and total angular  momentum unity. Such state is created in a small spatial volume of the order $ \hbar/\sqrt{q^2}=\hbar/(2M_p) \sim 0.1$ fm. At this point the system is a pair of  (colorless) proton and  antiproton each formed by three bare quarks (antiquarks). In agreement with parton picture, the quarks as partons have no structure. Therefore the Pauli FF vanishes and the Dirac FF is unity. It is then expected that $|G_E|=|G_M|=1$. The anomalous magnetic moment appears when the system starts to expand and to cool down. The current quarks (antiquarks) absorb gluons and transform into constituent quarks (antiquarks).  

2) The next step is the evolution of the created "hadron" pair with 'point-like' size. The residual energy turns into kinetic energy of motion, with relative velocity $2\beta =2\sqrt{ 1-4M_p^2/q_0^2}$. In the first stages, the strong chromo-electromagnetic field leads to an effective loss of color freedom of quarks and antiquarks. As a result of Fermi statistics, the identical (colorless) quarks ($uu$ in the proton and $dd$ in the neutron) are pushed out. The remaining quark (antiquark) of different flavor is attracted to one of the quarks at the surface, creating a compact diquark ($d u$ state). 

The neutral plasma acts on the distribution of the electric charge (not of the magnetic charge) as an additional suppression factor. Similarly and for the same reasons as in SL region, Eq. (\ref{eq:GESL}), also in TL region ($q^2>4M_p^2$) a suppression of the electric FF is expected :
\ba
|G_M(q^2)|&=&[1+(q^2-4M_p^2)/q_2^2]^{-2} \Theta(q^2-4M_p^2),\label{eq:GMTL}\\
|G_E(q^2)|&=&|G_M(q^2)|[1+(q^2-4M_p^2)/q_1^2]^{-1} \Theta(q^2-4M_p^2)\label{eq:GETL},
\ea
where we added the $\Theta$ function to avoid poles in the unphysical region. We replaced $Q^2\to q^2-4M_p^2$ in order to have the implicit normalization $|G_E|=|G_M|=1$ at the kinematical threshold. The magnetic FFs is defined by the distribution of the angular 
momentum (spin). It is expected that proton and neutron magnetic FFs are similar and that they are slowly varying functions of $Q^2$, as determined by quark counting rules. 

3) In the last stage, the long range color forces create a stable colorless state of proton and antiproton. Part of the initial energy is dissipated into the transformation of bare valence quarks and antiquarks into constituent quarks, giving rise to on-shell proton and antiproton. At this stage of evolution, they are separated by a distance $R$. The proton and antiproton tend to move apart with kinetic energy $T=\sqrt{q^2}-2M_p c^2$ which is counteracted by the opposite effect of the confinement energy: $(k_s/2)R^2$,  ($k_s$ is the elasticity constant due to confinement). 

For very small values of the velocity $\alpha\pi/\beta\simeq 1$ the final state interaction leads to the formation of a $N\bar N$ bound state, with size of the order of 100 fm. At larger distances, the inertial force exceeds the confinement force: proton and antiproton start moving, each one with velocity $\beta$. When the three quarks have receivent sufficient transferred momentum, they leave the interaction region. The spin unity of the intermediate virtual photon should manifests itself in dynamical  polarizations of the proton and the antiproton. 

$Q(t)$ (\ref{eq:eq14}) describes the time evolution of the distribution of negative and positive charges in the domain $\cal D$, which can be considered as a sphere of radius $tc$. Particular parametrizations of $Q(t)$ may reproduce the scaling behavior for $F(q^2)$, for example if ${\cal Q}(t)$ has some singularity in power $n$,  ${\cal Q}(t)\sim (t+t_0)^{-n}$. 

 At very large proton-antiproton distances the integral in Eq. (\ref{eq:eq14}) must vanish. 


\section{Comparison with existing data}

The world data on proton FFs together with the predictions of the model are shown in Fig. \ref{Fig:fig1f}. 


\begin{figure}[ht]
\begin{center}
\caption{World data on proton FFs as function of $q^2$. 
 {\bf Space-like region:}  $G_M$ data (blue circles), dipole function (blue line) from Eq. (\ref{eq:GMSL}); electric FF, $G_E$, from unpolarized measurements (red triangles) and  from polarization measurements (green stars). The green line is the model prediction from Eq. (\protect\ref{eq:GESL}). {\bf Time-like region:} world data for $|G_E|=|G_M|$ (various symbols for $q^2>4M_p^2$ and model prediction (black line) from Eq. (\protect\ref{eq:GMTL}).}
\label{Fig:fig1f} 
\end{center}
\end{figure}
In SL region, the magnetic FF, $G_M$ (blue circles) is well reproduced by the dipole function (\ref{eq:GMSL}) (blue line) normalized  to $G_M(0)=\mu $. The electric FF, $G_E$ from unpolarized measurements  (red triangles) also follows a dipole function normalized to $G_E(0)=1$.  $G_E$  (green stars) from polarized  measurements is well reproduced by the prediction of the present model (green line), Eq. (\ref{eq:GESL}), setting $q_1^2$=3.6 GeV$^2$. 

In the TL region, the individual determination of $G_E$ and $G_M$ has not been done, due to the lower luminosity achieved up to now. The experimental results, corresponding to the annihilation cross section, are given in terms of $|G_M|$, assuming $|G_E|=|G_M|$ or $|G_E|=0$. The data for  $|G_E|=|G_M|$ are shown for $q^2>4M_p^2$ in Fig. \ref{Fig:fig1f} (black circles) for various experiments, see Ref. \cite{Babar} and are well reproduced by the present model, following Eq. (\ref{eq:GMTL}) with $q_2^2$=1.2 GeV$^2$ (black line). 

The few existing data on the proton FF ratio in TL region are reported in Fig. \ref{Fig:ratiotl}.  The prediction of the present model is drawn as a solid line. The line is normalized at the kinematical threshold ($q^2=4M_p^2$) where $|G_E|=|G_M|$ (assumed $=1$). Due to the dispersion of the data, we do not try any fit, but we apply Eqs.(\ref{eq:GMTL}, \ref{eq:GETL}) with $q_1^2$=1.2 GeV$^2$.
\begin{figure}[ht]
\begin{center}
\caption{Proton electromagnetic FF ratio in TL region, as functions of $q^2$. The line is the prediction of the present model. Data are from Ref. \protect\cite{Ba94} (red squares), from Ref. \protect\cite{Babar} (black triangles), from Ref. \protect\cite{Ambrogiani} (green circle), and from Ref. \protect\cite{Fenice} 
(blue star). The shaded area shows the physical threshold.} 
\label{Fig:ratiotl} 
\end{center}
\end{figure}

The neutron and proton electric FFs in SL region are shown in Fig \ref{Fig:neutron}. A sample of the existing neutron data is shown (black triangles) (see Ref. \cite{Ri10} and Refs. therein). For comparison, the proton FF from polarization data  \protect\cite{Pu10} is shown (green, stars) and from Rosenbluth method \cite{An94} (red,triangles). The prediction is the solid, green line for the proton FF and the black, dashed line for the neutron following Eq. (\ref{eq:GESL}) with $q_1^2$=2.42 GeV$^2$. The vertical line indicates the limit of validity of the model for the neutron.

In TL region few data exist, based on only one experiment \cite{Fenice}. The model give a reasonable description taking into account that the data are not very precise, they have not been extracted with the point-like prescription at threshold, so that they partly include the Coulomb corrections.

\section{Discussion and conclusions}
 We have generalized the definition of FFs in all the kinematical range: the physical meaning of FFs in SL region is related to the charge and magnetic distributions in the Breit system, whereas in TL region it is related to the time evolution of the these distributions in CMS, as expressed in Eq. (\ref{eq:eq5b}). 

Our expressions and definition of FFs in TL region, have not been derived by analytical continuation from the SL region. For DIS, similar problems appear when ones tries to connect the annihilation channel which is related through crossing symmetry. Crossing symmetry holds in Born approximation. Quantum corrections associated to high order perturbation theory violate crossing relations, as demonstrated in QED \cite{Ku74}. Therefore discrepancies may appear in the fitting parameters, although the general trend of the parametrization is supposed to be similar and qualitatively reproduces the overall data.

The complex nature of FFs in TL region arises naturally from final state interaction, being larger in the threshold region and vanishing at large $q^2$, in agreement with the Phragm\`en-Lindel\"of theorem \cite{Ti39},  which applies to analytical functions of complex variables, and insures the equality of SL and TL FFs at large $|q^2|$. As FFs are real in the scattering channel, it has for consequence that the imaginary part must vanish in the annihilation channel at large $|q^2|$. 

The present qualitative picture is in agreement with a point-like nature of TL  FFs at threshold. Based on arguments on the absence of resummation of Coulomb corrections near threshold, in Ref. \cite{Ba09} it was shown that the extraction of FFs from the experimental cross sections is consistent with  $|G_E(q^2=4 M_p^2)|=|G_M(q^2=4 M_p^2)|=1$. 

In conclusion, we have developed an intuitive and qualitative model, which satisfies the known features of the data and the physical requirements. Moreover, we have given useful and simple parametrizations which reproduce satisfactorily the known behavior of all nucleon FFs data, in SL and TL regions at large $q^2$. 

\begin{figure}[ht]
\begin{center}
\caption{World data on neutron electric FF (black, triangle down), in SL region, as functions of $|q^2|$ from Ref. \protect\cite{Ri10} and Refs. therein. For comparison, the proton FF from polarization data  \protect\cite{Pu10} (green, stars) and from Rosenbluth method \cite{An94} (red,triangles) is shown. The present prediction is shown for the proton FF (solid, green line) and for the neutron (black, dashed line). The shaded area shows the region where the model does not apply for neutron ($q^2<2.43$ GeV$^2$) and the double shaded area shows the region where the model does not apply for proton ($q^2<1.21$ GeV$^2$).} 
\label{Fig:neutron} 
\end{center}
\end{figure}

\section{Acknowledgments}
One of us (E.A.K.) acknowledges IPN Orsay for warm hospitality. We acknwoledge Yu. M. Bystritskiy for careful reading. This careful work was partly supported by GDR No.3034 "Physique du Nucl\'eon" (France).






\bibliographystyle{elsarticle-num}
\bibliography{<your-bib-database>}







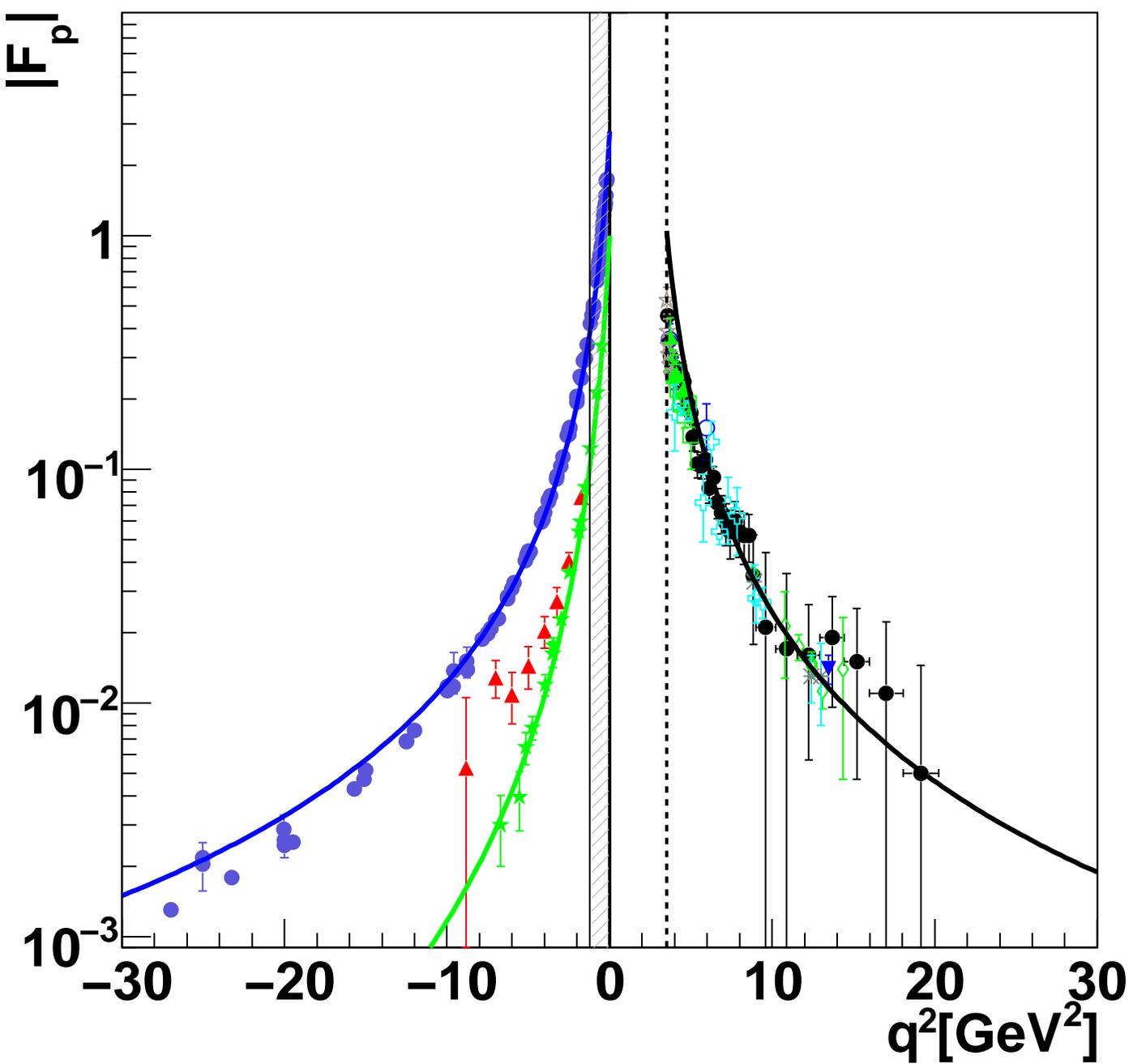
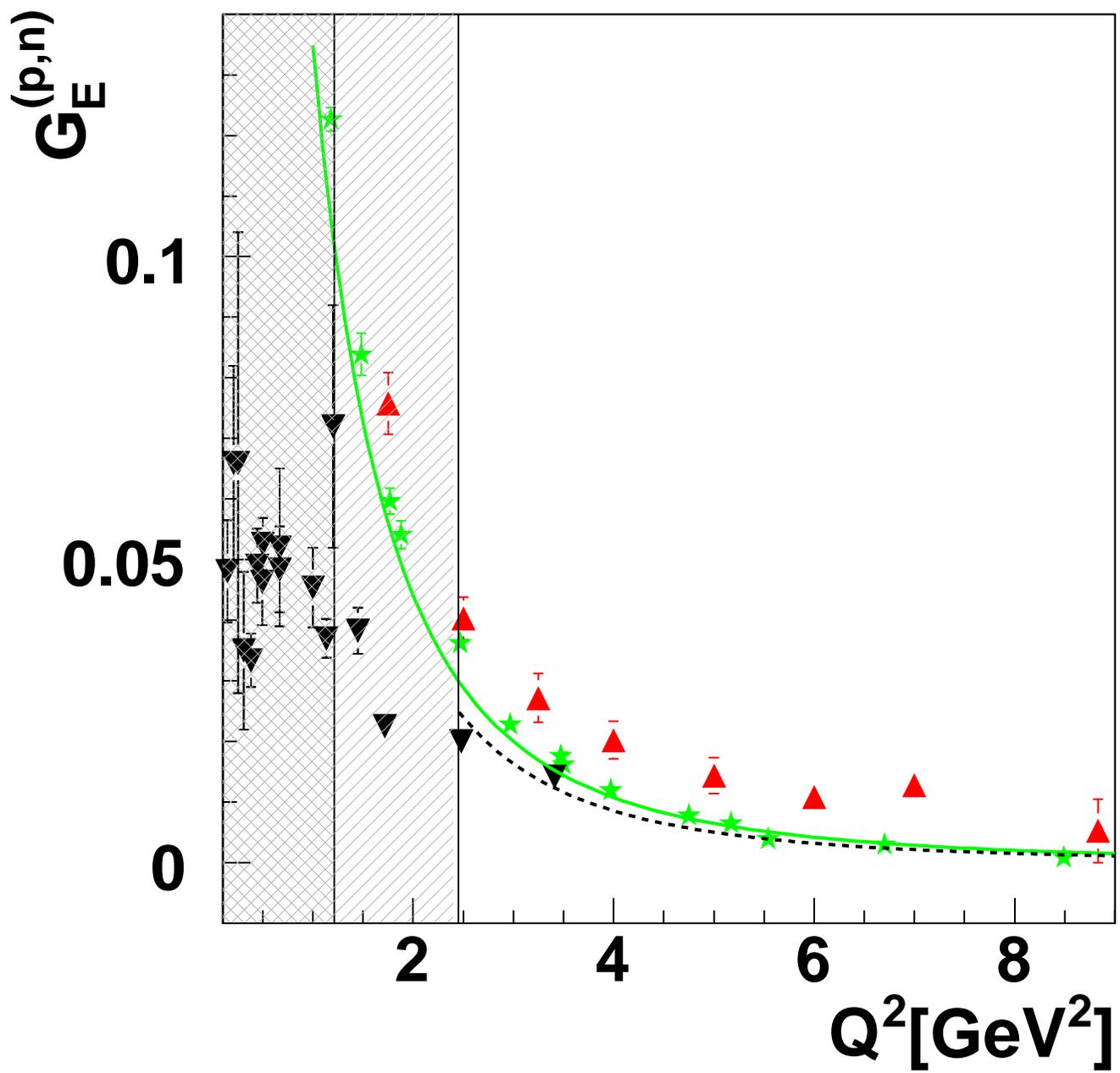
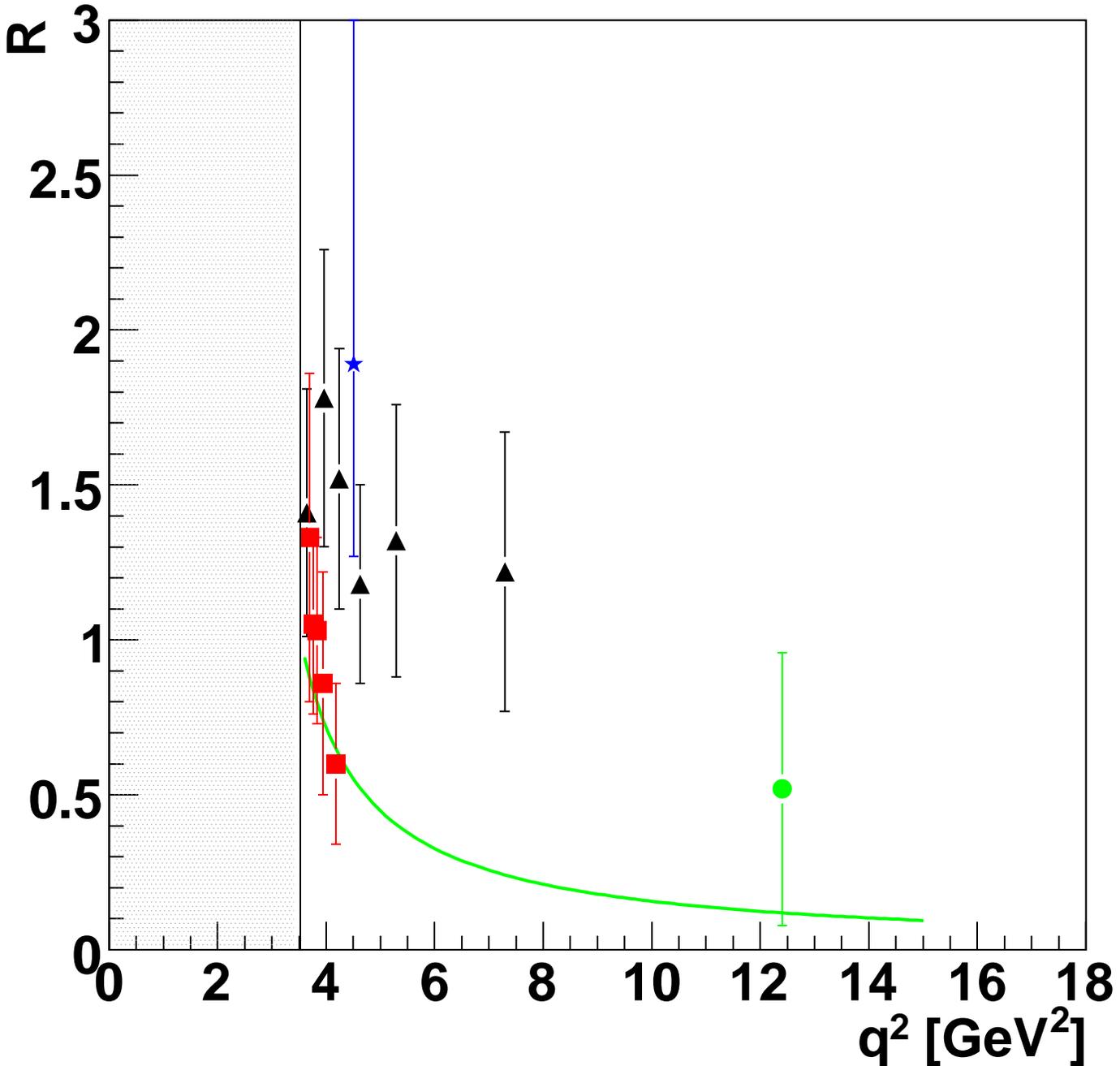
\end{document}